# A *limbus mundi* elucidation of habitability: the Goldilocks Edge

Ian von Hegner

Aarhus University

**Abstract** The habitable zone is the circumstellar region in which a terrestrial-mass planet with an atmosphere can sustain liquid water on its surface. However, despite the usefulness of this concept, it is being found to be increasingly limiting in a number of ways. The following is known: (i) Liquid water can exist on worlds for reasons unrelated to its specific distance from a star. (ii) Energy sources can exist for reasons unrelated to the distance to a star. Furthermore, the habitable zone is based on both astronomy – the distance and stellar energy – and chemistry – liquid water and the right temperature. However, these factors are only part of the consideration. Thus, discussions of habitability and the possibility for the emergence of life on a world must consider the evolutionary principles that govern life as well as the laws that govern stellar and planetary science. This is important because the following is also known: (iii) The temporal window for the emergence of life is within 600 million years. (iv) The Earth was an extreme environment overall in the period when this window existed. (v) The first life was necessarily fragile. Therefore, chemical evolution must have taken place in a relatively protected and restricted environment. Thus, rather than as in the Goldilocks zone, which focuses too narrowly on the world as a whole, this paper suggests that it is better to focus on a particular region and time period on a world, in which fitting conditions for habitability exist. Thus, the following is suggested: "The Goldilocks Edge is a spatial and temporal window on an astronomical body or planemo, where liquid solvents, SPONCH elements, and energy sources exist." Furthermore, since the mere presence of these do not in themselves necessarily lead to the emergence of life, this possibility only arises when these interact. Thus, it will also be suggested that the following can exist within the Goldilocks Edge: "The great prebiotic spot is an environmentally relaxed and semi-shielded area on an otherwise extreme world, wherein conditions advantageous for chemical evolution exist." Such Goldilocks Edges, and the great prebiotic spots within them, can collectively represent the full distribution of possibilities for life in a solar system.

## 1. Introduction

The concept of a habitable zone, introduced in the 1950s, has become particularly influential; it has shaped the search for specific exoplanets in astronomy and impacted astrobiology as well. The term refers to the region around a star where liquid water can exist on a planet; the more informal term "Goldilocks zone" was suggested by NASA scientists in the 1970s. The latter is an analogy to the story of "The Three Bears," wherein a child named Goldilocks tastes three different bowls of porridge, eventually ignoring the ones deemed too extreme, too hot or too cold, and preferring the one in the middle, which has the right temperature. This is similar to the concept of a region around a particular star where the temperature is "just right" for the liquid phase of water to occur.

The habitable zone is thus very much based both on astronomy—because the boundaries of such a zone are based on the distance from the star and the resulting stellar energy it receives—and on chemistry, with its focus on liquid water and the requirements for a particular temperature. However, although a liquid solvent and the provision of energy are necessary conditions for life, these are nevertheless still only a part of the total picture.

Astrobiology is the offspring of both the Copernican and Darwinian revolutions, the two events which profoundly changed the perception of the position of the Earth in the Universe and the position of a particular species in nature.

Thus, discussions of habitability and the possibility for the emergence of life on a world must take into account the evolutionary principles that govern life as well as the laws that govern stellar and planetary science. Evolution concerns not merely life as we know it. Just as the principles of say thermodynamics are conjectured to be universal, it can also be conjectured, that the evolutionary principles that govern chemical and biological evolution are universal.

If there is life elsewhere in the universe, then it may be quite different in terms of phenotype, but this would be a consequence of the natural selection and mutation which are both integral parts of Darwinian evolution, which by itself is a non-random process. The limiting influence of evolution therefore defines the range of life bearing worlds, and predictions made in accordance with this process look for habitable worlds alongside the known parameters of planetary science.

There has recently been significant interest in the moons Europa and Enceladus within the discipline of astrobiology, as they appear to possess the right conditions necessary for life. It has also become apparent that the planet Mars once had adequate conditions for the development of life, and may still have the capacity to harbor extremophile analogues. If the conjecture is granted, that there has been or that there is life on these worlds, then there could be four worlds on which life has emerged. They therefore represent the possible distribution of life in the Solar system.

Whether this is the case, and whether this is possible, are questions that are imperative in the search for extra-terrestrial life in the galaxy. The Earth is in a habitable zone; Mars is thought to be either inside or just outside this zone, depending on the model used; but Europa and Enceladus both lie in the circumplanetary zone. However, the assertion that life mainly exist on Earth-like planets in what is called the habitable zone is still only an assertion. It may instead be the case that the habitable zone represents an extreme value, or the tail of a curve, while inhabited worlds outside this zone represent the mean value on a Bell curve.

This is an important consideration because it is now known that there are more planets than stars in the galaxy, and more planets outside habitable zones than in them. There are also presumably far more moons than planets in the galaxy. Thus, there may be a great number of inhabited worlds outside habitable zones.

The concept of a habitable zone is very much a bottom up approach, with a focus on the conditions necessary for liquid water and temperature. However, the possibility of the right conditions for life on a world is not the sole indicator that life will actually develop on that world. Even the possibility for chemical evolution on a world is not the sole indicator for life on that world (phenomena such as RNA or autonomous cycles are impressive, but they are not life). After all, only actual biology is truly indicative of life.

The emergence of life and its continuous existence is not merely centered around the presence of liquid water and temperature. Chemical and biological evolution are active processes. Thus, this approach must be supplemented by a top down approach founded on the principles of chemical and biological evolution, since evolution can be used to retrogressively produce information concerning the active conditions which must have been present for the emergence of life.

A synthesis of planetary science and evolution will enable a fertile focus on the world itself, on the demands life put forward, rather than keeping focus on the distance of a world to its star and the relationship therein. Thus, the concept of a '*limbus mundi*' or 'world's edge' may provide an answer to the question of inhabited worlds.

## 2. Discussion

This paper is organized as follows: Section 2.1 will clarify the conditions required for habitability, of which there exist many examples can exist outside the habitable zone. Section 2.2 will clarify some of the evolutionary conditions that hold for life itself, and how these can be used to retrogressively demonstrate the demands life require for its emergence on a world. Section 2.3 will conjecture how the prebiotic conditions must be before the emergence of life on a given world. Section 3 will discuss the proposition that instead of solely focusing on a specific concept such as the habitable zone, instead focus should be on a more general concept such as the Goldilocks Edge. Section 4 will summarize the results of this investigation and discuss the implications in the search for habitable worlds.

### 2.1. The habitable zone

The Earth is so far the only world on which life is known to exist. This singular reference point means that defining the essential requirements for life is challenging. The focus has generally been around why the location of a particular world in the Solar System makes it fit for life. It is evident that all life on Earth requires liquid water, which is conjectured to be essential for all life and thus the most important feature of a habitable planet. The position of the Earth within the Solar systems allows for the existence of liquid water on its surface.

The term "habitable zone" was first introduced in 1959 to refer to the region surrounding a star where it is possible for liquid water to exist. This term was introduced within the context of habitability of planets in other solar systems, as well as concerning the possibility of life on such worlds [Huang, 1959]. Thus, the habitable zone is defined as the circumstellar region in which a terrestrial-mass planet with a $CO_2$–$H_2O$–$N_2$ atmosphere can sustain liquid water on its surface [Kasting et al., 1993; Kopparapu et al., 2013], or more precisely, where liquid water can be maintained on the surface at a particular instant in time.

Estimations of the extent of the habitable zone vary for several reasons. The long-term estimation for the habitable zone of the Sun extends from 0.95 to 1.37 AU [Kasting et al., 1993]. Newer models have recalculated the zone as extending from 0.99 to 1.70 AU [Kopparapu et al. 2013] while others have estimated the inner edge to be even closer to the star at 0.93 AU [Wolf and Toon, 2014].

Furthermore, as the stellar luminosity of stars increases over time, the habitable zone evolves outwards as the boundary of this zone is not static [Hart, 1978]. Thus, a "continuously changing habitable zone," which can be defined as the region around a star where the conditions that support the presence of liquid water are present for a specified period, exists. A conservative estimate for the width of the zone in the Solar System is 0.95 to 1.15 AU [Kasting et al., 1993].

However, despite the usefulness of the concept of a habitable zone, it has increasingly been found limiting for a number of reasons. The idea that habitability requires an Earth-like planet with a $CO_2$–$H_2O$–$N_2$ atmosphere and the presence of liquid water on its surface seems unnecessarily restrictive. There is now ample evidence for the presence of liquid water outside the circumstellar habitable zone.

Mars orbits outside the habitable zone [as defined by Kasting et al., 1993]. During the Noachian era, the atmosphere on Mars was probably denser than it is at present, and the planet may have therefore been relatively warm and wet, or cold with ice that is melting [Fastook and Head, 2015]. The presence of liquid water on the surface of the planet at some point in the past is indicated by several lines of evidence, such as the existence of ancient, water-eroded structures and the presence of weathered exhumed phyllosilicates [Carter et al., 2010]. It is even possible that the northern plains were covered with an ocean [Brandenburg, 1987].

The classical habitable zone is concerned only with the occurrence of liquid water on the surface of a world's surface, but it is now accepted that liquid water can exist beneath the surface and indeed, on the surface of worlds orbiting outside the habitable zone. While liquid water was probably present on the surface of Mars at some time in the past, it has become apparent that liquid water may still exist in the subsurface [McEwen et al., 2011]. Europa is one of the four Galilean moons of Jupiter. It has an outer crust of solid ice and is thought to have a deep global ocean of liquid water beneath this surface [Kivelson et al., 2000] with a depth of approximately 100 km [Chyba and Phillips, 2002]. Likewise, Enceladus is a moon orbiting Saturn that is also thought to have a global subsurface liquid ocean beneath its frozen surface [Thomas et al., 2016].

The presence of liquid water is possible for several reasons. The first is that it can exist due to the distance between a star and the atmosphere of a planet or moon, as on Earth. It can also occur due to tidal heating—as on Europa—where complex gravitational interactions lead to the moon moving in a noncircular orbit around Jupiter, resulting in a tidal bulge that fluctuates in accordance with the orbit around the planet. This flexing causes tidal heating, allowing the underlying ocean to remain in a liquid state. The icy shell at the surface of the moon insulates the liquid water underneath, trapping the heat [Barr and Showman, 2009]. Liquid water can also be present because of other mechanisms that produce heat, such as the subdominant tidal force caused by the non-zero axial tilt of Europa; this generates large Rossby waves and produces energy that is at a minimum 2000 times greater than that created by tidal heating [Tyler, 2008].

Liquid water, or at least pockets containing liquid water, can also form due to the thermal energy generated by a world itself. Hydrothermal vents that are observed at the bottom of the ocean are well known on the Earth, where hot fluid is present beneath the ocean floor, and are caused by underlying magma that is close to the surface [Colín-García et al., 2016]. Liquid water could also exist on a so called rogue planet in interstellar space if it had both a high-pressure heat-trapping hydrogen atmosphere and a geothermal heat flux, ensuring that although the effective temperature of the world is in the order of 30 K, the temperature at the surface is still capable of carrying liquid water [Stevenson, 1999]. A subglacial liquid ocean could presumably be maintained on a rogue planet if thick layers of thermally insulating water ice and frozen gas were present together with a geothermal heat flux provided by both the decay of radioisotopes and heat originating from the formation of the world [Abbot and Switzer, 2011]. Thus, the presence of liquid water is possible on a multitude of worlds under many different conditions.

**2.2. The distribution of life**

A process called OoL (Origins of Life), or chemical evolution, is the natural process or transition by which life can emerge from the synthesis of the simplest organic building blocks sourced from inorganic substrates, that eventually leads to self-sustaining replicating systems [Scharf et al., 2015]. While the details are still under debate and much remains to be elucidated, it is clear that prebiotic processes must have eventually led to the emergence of the first fully autonomous cell, which is known with a high certainty to have existed 3.5 billion years ago [Schopf et al., 2007]. Biological evolution itself began with this singular event, when chemical evolution locked it on to this very first autonomous cell. Evolutionary theory illuminates two facts that are relevant to this discussion.

The Full House model was put forward by Gould (1996) in order to explain that although the phenomenon of increased complexity is evident in the tree of life, complex life arises only as a side consequence of a physico-chemical constrained starting point, and actually represents only a relatively minor phenomenon. This statistical model shows that complexity represents only the small right-hand tail on the bell curve for life, with a constant mode at bacterial complexity [Gould, 1996]. The Full House model suggests that due to the constraints imposed onto the origins of life from chemical evolution and the physical principles of self-organization, the first life form necessarily came into existence at the lower limit of life's conceivable and preservable complexity. Thus, the first life form is imposed to begin with the simplest starting point right next to a lead wall of complexity. This physico-chemical lower limit can be designated the 'left wall' in an architecture of minimal complexity [Gould, 1996].

Since biological evolution began with the simplest possible functional life, the first life cannot have been an extremophile despite the fact that the environment on the early Earth was extreme compared to that of present Earth. Thermophiles are probably the oldest extremophiles on this planet, and may have been the first adaptation that life underwent. A model reconstruction of ancient RNA indicates that LUCA, the last universal common ancestor, was either a thermophile or hyperthermophile [Gaucher et al., 2010]. However, since the first life was at the lower limit of life's preservable complexity, extremophiles cannot have been the first life as they are not the simplest conceivable organism. Biological evolution is a stepwise process, in which adaptations take place over time. Since extremophiles are more complex than the first life, there necessarily must have been an intermediary duration between the first life and the extremophiles, during which natural selection allowed the adaptation of organisms to their immediate environments [von Hegner, 2019].

Thus, two facts need to be addressed: the first is that the first life was fragile compared to later life and the second is that life requires time to adapt to the surrounding environment. This means that the life emerging in the hostile environment of the early Earth necessarily must have formed and evolved in a relaxed protected area.

**2.3. The Great prebiotic spot**

Several lines of evidence point to the emergence of life on Earth between 4.1-3.5 billion years ago [Bell et al., 2015], and it seems that the transition from chemistry to biology was not a single event, but was instead a gradual series of steps of increasing complexity.

The available evidence thus points to a window of approximately 600 million years that was available for the emergence of life on the Earth. It is not this window of time in itself that is interesting here however, what is interesting is that the main part of this window occurred during a very violent period in the evolution of both the Solar system and the Earth.

The Earth today has favorable conditions in which life can thrive, but this was not the case at this time in the Earth's history. The Earth of between 4.1 and 3.5 billion years ago was very different than it is today. Earth was formed via accretion from the solar nebula ±4.5 billion years ago, and the Hadean eon describes the geological period from the end of accretion until 3.8 billion years ago [Coenraads and Koivula, 2007]. At this time there was only a thin atmosphere made up of hydrogen and helium, and the solar wind removed much of any atmospheric gases produced. An atmosphere could only begin to accumulate when a core-generated magnetic field formed that could protect the gases released during volcanism.

In this phase of its development the Earth was violent, with constant bombardment from impactors. A large part of the Earth is theorized to have been almost destroyed by a major impact with a body the size of Mars within the first 100 million years of its existence, leading to the formation of the Moon [Canup and Asphaug, 2001]. Eventually the molten surface reformed into a solid crust, and the thick atmosphere allowed for the condensation of water even at such high temperatures under the high atmospheric pressure. Thus, the steam escaping from the crust and gases released during the early volcanism built up the prebiotic atmosphere, while heavy rain fell, forming the emerging and boiling oceans [Coenraads and Koivula, 2007]. The bombardment of the Earth continued, ending in the late Heavy Bombardment period which took place between 4.1 and 3.8 billion years ago, although these dates are still debated [Lowe and Byerly, 2018]. This was the most intense period of bombardment.

The Archean eon occurred from 3.8 to 2.5 billion years ago and was likewise a very active period. The surface was hot and the continents had presumably not yet formed. Intense plate tectonic activity occurred at the beginning of this period, melted and then reforming the surface. Near the end of this eon, tectonic activity decreased to a level almost equivalent to that of present-day Earth [Coenraads and Koivula, 2007]. Violent volcanism had been present almost continuously when the first continental crust began to form, but the temperature had decreased, and by the beginning of the Archean eon global oceans, which may have begun to form by 4.2 billion years ago, covered most of the planetary surface [Cavosie et al., 2005].

The atmosphere was enriched with water vapor, carbon oxides, nitrous oxides, hydrogen sulfide, methane, ammonia, and other noxious gases. However, free oxygen is thought to have been nearly absent during this time. Two processes are responsible for introducing oxygen into the atmosphere. One began during the Hadean via photochemical dissociation, accounting for 2 % of the present-day oxygen, which was sufficient for the formation of enough ozone to create a protective barrier against ultraviolet radiation from the Sun. The second source of oxygen was via the activity of photosynthetic organisms, which fundamentally changed the atmosphere of the Earth during the Archean [Coenraads and Koivula, 2007].

There is still much debate regarding the details of these first two eons. However, the Earth was clearly overall an extreme environment during the period when this window existed, and yet life still managed to emerge during the transition between the Hadean and Archean Eons which has been puzzling. Because the first life must necessarily have been fragile, as discussed in section 2.2, and must necessarily have required a certain duration of time to adapt into extremophiles better suited to handle that environment.

Although chemical evolution can be driven by highly energetic processes and high temperatures, stages of its development must have been protected. For example, RNA is a chemically fragile biopolymer which is unlikely to survive in extreme conditions, and cell membranes would likewise have been too fragile to exist in such an environment. Therefore, unless panspermia, whereby genetic material is either assumed to emerge on and be transported by comets to planets [Hoyle and Wickramasinghe, 1986] or be transported by asteroids or meteorites to and from different planets or solar systems [Belbruno et al., 2012] is assumed to be the mechanism whereby such material can occur on the surface during such harsh conditions, then these facts regarding the emergence of fragile life on a world undergoing extreme conditions must be accepted.

If the harsh environment of the early Earth is taken into consideration, these facts are puzzling. However, this changes if we imagine that life emerged in a restricted environment that was different from general conditions on the planet—that there was a spatial and temporal window fit for chemical evolution somewhere in the extreme environment.

Thus, given these considerations, the existence of the great prebiotic spot is proposed. It can be defined as follows: *The great prebiotic spot is an environmentally relaxed and semi-shielded area on an otherwise extreme world, wherein conditions advantageous for chemical evolution exist.*

The spot is a persistent region where liquid solvents, SPONCH elements, and energy sources interact dynamically with each other. Chemical reactions would be able to take place in such a region, experiencing selective pressure, and chemical replication would occur in increasingly complex forms that eventually may lead to the variety of organizational phenomena associated with life.

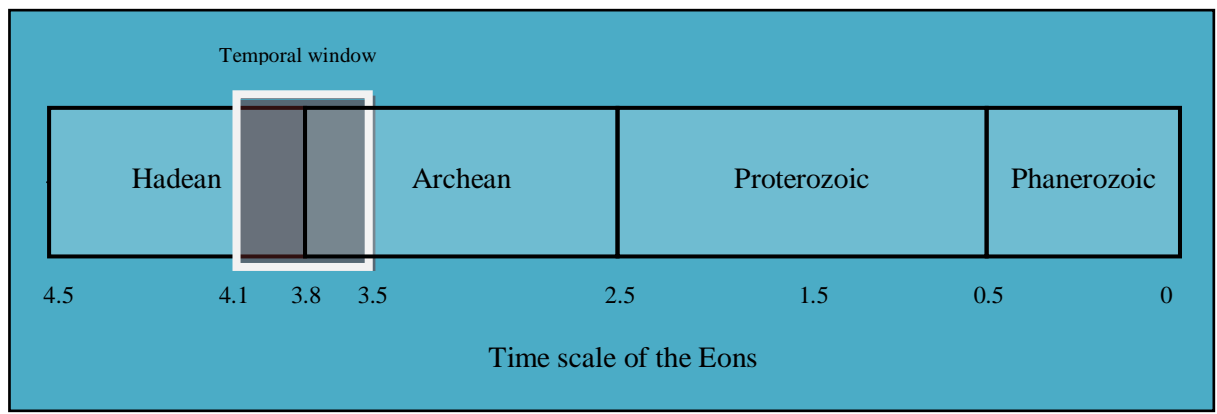

**Figure 1.** Time scale of the geologic Eons of the Earth in billions of years. The location of the temporal window is located in the Hadean and Archean Eons.

Quantifying the great prebiotic spot is not easy due to its very nature. The great prebiotic spot have long ago disappeared and biological evolution has taken its place. Furthermore some factors that would have been integral to such a phenomenon would by their very nature be challenging to quantify. However, a number of physical and chemical restrictions which must have been in place can be deduced and used to guide a qualitative discussion. These restrictions can be formulated as the following inequalities:

$$T_{window} \leq T_{600 \text{ million years}} \quad (a)$$

$$S_{world} > S_{prebiotic\ volume} > S_{0.4\ \mu m^3} \quad (b)$$

$$C_{RP} \neq C_Q \quad (c)$$

$$E_{external} > E_{internal} \quad (d)$$

where $C_{RP}$ is the chemical reactions, $C_Q$ is chemical reactions multiplied by k, k is the variability factor or the deterministic reduction coefficient K, $E_{external}$ is the increase in the entropy of the surrounding environment, and $E_{internal}$ is the reduction in entropy in the great prebiotic spot.

**Temporal window T.** The great prebiotic spot has a temporal window T, which is less than or equal to 600 million years which was available for the emergence of life, as suggested by evidence that was mentioned previously, as detailed in Fig. 1. Only one sample is as yet available for the emergence of life, and as no examples have as yet been located on other worlds or emerged in a laboratory, it is not yet known how much time is required for its emergence. It is possible that life can emerge much faster under more ideal conditions than those apparent on the early Earth.

There may have been many such spots on the Earth, of different size and with varying degrees of favorable conditions for chemical evolution. The period of time may be more important than the size of the great prebiotic spot, or an area of significant size containing an abundance of organic molecules may be more important. It is not yet known. But as yet the temporal window for the Earth must be used as a restriction.

**Spatial size S.** The minimum spatial size S of such a great prebiotic spot is not easy to estimate because of the lack of data. Apart from being sufficient for the presence of a liquid solvent and SPONCH elements, such a region would also have to be large enough for repeated chemical reactions to take place and have enough space for selective pressure to occur among the products of these reactions. Darwin (1871) suggested in a brief afterthought that life could have begun in "some warm little pond with all sort of ammonia and phosphoric salts,—light, heat, electricity present" [Darwin, 1887], while Haldane (1929) suggested that life emerged from molecules that "accumulated till the primitive oceans reached the consistency of hot dilute soup" [Haldane, 1929].

A large ocean could appear having sufficient materials. However, not only does the concentration of monomers in a large ocean appears to be too low for polymerization to take place, but it would also have been unprotected from the harsh environment of the Earth at that time. However, biological insight can be used to find the minimum size of a potential great prebiotic spot. The first life was the simplest possible life. The average size of a bacteria is between $\sim 0.4 - 3$ $\mu m^3$ [Levin and Angert, 2015]. Bacteria of both larger and smaller size do exist, but are generally evolved from bacteria of average size, representing only the small right tail of a bell curve, whereas the bacterial mode itself has remained constant in size. The size of the first life was therefore presumably close to this size, and a great prebiotic spot therefore cannot have been smaller than this. Thus, a pragmatic lower limit for a great prebiotic spot can be set at 0.4 $\mu m^3$. This is very likely significantly lower than any actual formation, since also the above suggested pond presumably would be too small to accommodate the processes which lead to the emergence of the first cell, as there would probably be insufficient material present in such a small space. However, this reasoning provides a boundary for the minimum size of a great prebiotic spot.

The upper size of a great prebiotic spot has to be smaller than the body on which the life has developed; how much smaller it was would probably depend on planetary and chemical conditions. Thus, there would probably be an apparent difference in the size of great prebiotic spots on early Earth and Europa.

**The processes $C_{RP}$ and $R_O$.** The great prebiotic spot is a persistent region where there is an increased probability that the transition from abiotic to the biotic reactions would occur. This probability is important to mention. The chemical reactions $C_{RP}$ are deterministic *per se*; they have directionality. How they will happen and what will happen is known. The rate of a chemical reaction is given in units of concentration and time such as M/s, where M is molarity (moles per liter) and s is time in seconds; the units will change depending on the overall order of the chemical reaction.

However, chemical reactions do not automatically lead to life. They are not chemical evolution $C_Q$. Although evolution is a non-random process, it is still bound by contingent events and selective pressures. This is one of the characteristics of chemical and biological evolution, which truly sets evolution apart from purely physical and chemical phenomena. Thus, the path from chemical reaction to chemical evolution is through a dynamic and non subjective factor k. This factor varies with contingent events and selective pressures, which could therefore be designated the variability factor or the deterministic reduction coefficient K.

Quantifying contingent events may be challenging because of their random and singular occurrences. Selective pressures take place continuously in chemical and biological evolution, although quantifying these in a natural setting may also be challenging. However, the factor may be given by the following heuristic formula:

$$k = ab(N_1 - N_2) \qquad (1)$$

where a is contingency pressure, b is selective pressure, and a < b. $N_1 = 1$ and $N_2 = 1 + s$ are natural variants in which s is the fitness advantage s > 0 or disadvantage s < 0 of $N_2$.

Homochirality is a good example here. If molecules composed of the left-handed form of amino acids (L-amino acids) possess even modest advantage over right-handed amino acids in a given environment, then selective pressure will select the former. Furthermore, although the amino acids found in life on Earth are in the left-handed form, this could be due to a contingent event. If for instance Europa is in fact teeming with life, there is no reason why its amino acids could not be right-handed, if a contingent event has pushed the selective pressure in that direction.

**Environments $E_{internal}$ and $E_{external}$.** The great prebiotic spot is a system in which an internal environment that is different from the surrounding external planetary environment exists. From a purely material angle, it represents a favorable concentration of prebiotic materials, i.e., the intermediary stage between abiotic and biotic factors. However, from an energetic angle it represents the presence of thermodynamic disequilibrium and simultaneously a temperature appropriate for chemical bonding. Thus, it is a system far from thermodynamic equilibrium, where an overall reduction in entropy $E_{internal}$ is occurring, via the displacement of entropy, to the surrounding environment $E_{external}$.

It is well known that order can arise in such systems. A dissipative structure is an open thermodynamic system that operates far from equilibrium and is characterized by a spontaneous structural and functional order and a low value of entropy [Prigogine and Lefever, 1968]. Prigogine demonstrated that the order within an open thermodynamic system will increase as energy flows through it. Examples of such systems in nature are hurricanes, fire whirls and of course the biological cell itself. The biological cell is an open system far from thermodynamic equilibrium; it maintains order by take in free energy from the surrounding environment and returns entropy in the form of heat and waste products [Toussaint et al., 1992].

Thus, the great prebiotic spot is a persistent system in which aggregations of material obtain order, complexify, transform, and accumulates information. The probability of the emergence of life is greater in such an internal dynamic environment than in the surrounding environment.

Can such great prebiotic spots exist outside the habitable zone? The great prebiotic spot is a localized region that is advantageous for chemical evolution. As it is not therefore a world as a whole, then it can be extrapolated that it is possible for such a region to exist on bodies that are otherwise considered extreme. It is known that life could emerge on an overall extreme world such as the Earth in ancient times. It is known that environments with liquid water exist elsewhere in the Solar system. It is known that SPONCH elements are widely distributed in the Solar system. There is evidence that SPONCH elements are present in the ocean of Enceladus, as gaseous $H_2O$, $CO_2$, $CH_4$ and $NH_3$ were detected by the Cassini spacecraft [Waite et al., 2009].

What is necessary in an environment is energetic disequilibria, meaning, there is energy to be harvested, meaning, that fundamentally there must be a presence of adequate sources and sinks for the free energy. It is well known that the energy used to drive chemical processes can be derived from sources other than the sun. For example, heat from geothermal processes is a standard energy source that can be used for chemical reactions [Muller and Schulze-Makuch, 2006].

It thus seems possible that great prebiotic spots could exist on Europa and Enceladus, which are far from the habitable zone. They could have existed on ancient Mars. They may even exist on the moon Titan. Thus, it appears to be the case that such great prebiotic spots can exist on many different worlds for varying time periods, and their size can depend on conditions which vary from world to world. If life has not emerged on Europa or Enceladus already, then it is possible that a great prebiotic spot may be developing there now.

The existence of such spots on different worlds does not necessarily mean that life will emerge there; only that there is an increased possibility of the emergence of life there. It means, in other words, that there is a Goldilocks Edge there.

**3. The Goldilocks Edge**

It is now known that liquid water can exist for a multitude of reasons, as discussed in section 2.1. Liquid solvent, SPONCH elements, and energy sources all occur outside the habitable zone. Energy that can drive chemical reactions can obviously be derived from solar energy. However, a variety of other energy sources are available on planetary bodies.

Heat is a well-known (endothermic) source of energy. Thermal energy can also be an energy source for chemical evolution, provided there is a heat sink as any gain in free energy requires the presence of a thermal gradient [Muller and Schulze-Makuch, 2006]. Heat can be provided by tidal heating, as is the case of Europa, where complex gravitational interactions maintain Europa in an eccentric orbit around Jupiter that results in regular fluctuations in the tidal bulge on the body of Europa. This flexing causes the ocean to heat up [Barr and Showman, 2009]. There may also be a separate heating method, whereby a subdominant tidal force that is caused by obliquity is able to generate large waves, which provides a heat source on Europa that is two thousand times larger than that from the previously mentioned tidal forces [Tyler, 2008]. Heat can also be derived from the world itself. The hydrothermal vents that occur around plate margins at the oceanic floor of the Earth are well known, some with relatively high temperatures. In these phenomena the hot fluids flow from beneath the surface to the ocean floor, caused by the heat source from the underlying magma [Colín-García et al., 2016].

A source of energy could also be based on the hydrogen produced by the reactions of rock with water. One microbial community of methanogens is known to consume the $H_2$ produced by the serpentinization of olivine in the rocks of the Earth [Stevens and McKinley, 1995]. Energy is also available from naturally occurring radioactive energy sources, where for instance actinides can act as fissile fuels in nuclear reactors whose power production can assist in chemical evolution [Adam, 2007]. Another source of energy could be derived from the redox reaction produced by radioactive decay. Some sulfur-reducing bacteria do in fact consume the hydrogen and sulfate produced by radioactive decay, on the Earth [Lin et al., 2006].

Thus, it seems clear that the emergence of chemical reactions that could be used for chemical evolution do not require a certain distance to a star *per se*. The right conditions for life are not the same as a simple binary choice, where one like Goldilocks select from sets of three items, discarding the worlds deemed too hot, too cold, too wild, too quiet, and focusing only on the worlds with the overall right conditions. It can be suggested that a world as a whole does not require suitable conditions for life; it only needs to possess a restricted possibility for a relaxed environment.

The focus can therefore be shifted from the relationship between star and planet to a world itself, and the concept of a generic Goldilocks Edge can be suggested: *The Goldilocks Edge is a spatial and temporal window on an astronomical body or planemo, wherein liquid solvents, SPONCH elements, and energy sources exist.*

An astronomical body is a tightly-bound structure kept in relative balance by gravity; it includes asteroids, comets (with reference to their solid frozen nucleus), moons, and planets. A planemo or planetary-mass object is defined as a round nonfusor while a planet is a planemo orbiting a fusor. A fusor is an object capable of core fusion during its lifetime, e.g., brown dwarfs and stars. Thus, the large round moons in a Solar System are planemos, but not planets, and free-floating objects such as rogue planets are planemos [Basri, 2003].

The requirement of a liquid solvent for the hosting of chemical interactions is conjectured to be a universal necessity, and since all known life on Earth uses water as a solvent, water is assumed to be the universal solvent. There are good reasons for this assumption. Hydrogen is the most abundant element in the Universe, while oxygen is the most abundant element on Earth [Mottl et al., 2007]. Thus, liquid water exists in large quantities on Earth. Furthermore, the hydrogen bonding capacity and the polarity of water allows for the dissolution of numerous substances [Pohorille and Pratt, 2012]. However, although the concept that water is the most probable solvent on any inhabited world is reasonable, the idea that other solvents can be utilized elsewhere cannot be rejected because of contingent events. Thus, alternative solvents such as ammonia, methane, ethane, nitrogen, or sulfuric acid could dominate on some worlds. Thus, the term liquid solvent will be used here.

All terrestrial life shares a common chemical composition, using the same essential elements to form biomolecules. Thus, the elements carbon, hydrogen, nitrogen, oxygen, sulfur, and phosphorus, which have been designated CHNOPS, or the more linguistically appealing, SPONCH, are considered to be the six elements most essential to life. The term SPONCHSi* should perhaps be favored instead, where the star designates, that it is known that at least 77 elements interact with terrestrial life [Committee on the Limits of Organic Life in Planetary Systems, 2007], and that silicon-based life appears to be a possible alternative to carbon-based life. However, SPONCH elements are known to form the basis of the biopolymers used by life, such as DNA and lipids, while no organism has been found as yet where silicon is used as the core for the formation of such structures. Thus, for simplicity only SPONCH elements will be considered here.

The Goldilocks Edge can be expressed using:

$$G_{passive} = \frac{L_S + Z + E}{V_{SC}} \qquad (2)$$

or

$$G_{passive} = \frac{L_S + Z + E}{V_{RSS}} \qquad (3)$$

where $L_S$ is the liquid solvent, Z is the SPONCH elements, E is the energy source, and $V_{SC}$ or $V_{RSS}$ is the volume of the restricted Goldilocks Edge that varies whether this edge include the surface on a world, or takes place under the surface of a world.

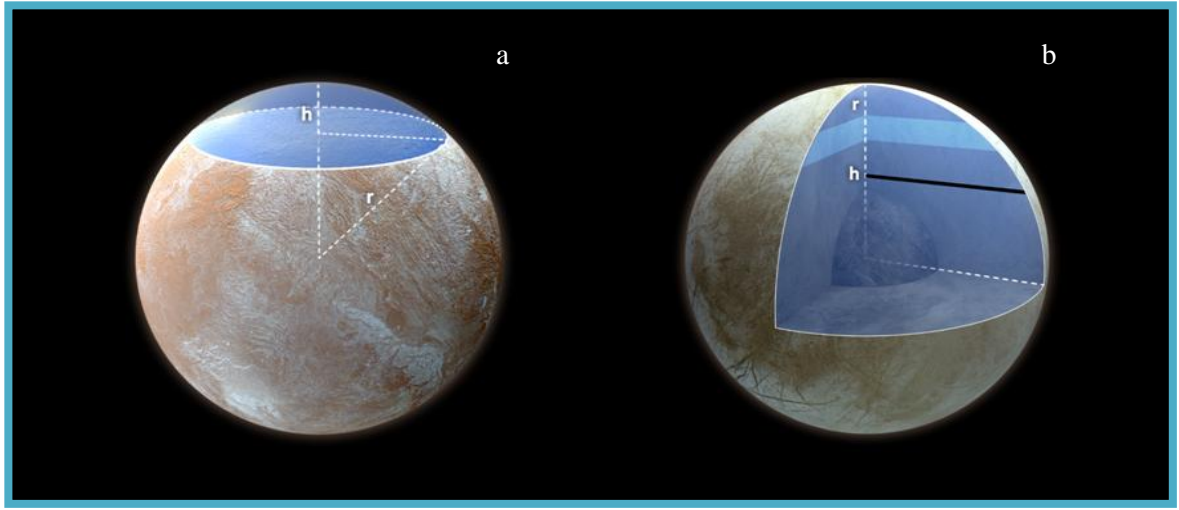

**Figure 2.** The Goldilocks Edge. (a) For early Earth-like worlds, a good approximation of the edge can be expressed as a spherical cap. (b) For worlds such as Europa, a good approximation of the edge can be expressed using a restricted spherical segment. Credits: (a) adapted from NASA/Goddard Space Flight Center/Francis Reddy, and (b) adapted from NASA/JPL/DLR.

Thus, for early Earth-like worlds the volume can be expressed as a spherical cap to good approximation, as in the region of a sphere which lies above a given plane, as seen in Fig. 2a. The derivation of this equation is well known (see Weisstein) and can be derived from the equation for a spherical segment:

$$V_{SS} = \frac{1}{6} \pi h (3r_1^2 + 3r_2^2 + h^2). \tag{4}$$

If $r_2 = 0$, and the Pythagorean theorem is applied, then we obtain:

$$V_{SC} = \frac{1}{3} \pi h^2 (3r - h). \tag{5}$$

For worlds such as Europa, the volume can to good approximation be expressed using a restricted spherical segment. A spherical segment is the solid defined by cutting a sphere using a pair of parallel planes (see Weisstein). But rather than using the well known equation for this, it will be more fitting for the purpose of this paper to derive an equation for the volume of a restricted spherical segment $V_{RSS}$. Thus, let a sphere of radius r be cut by a plane h units above the spheres equator, where h < r, as seen in Fig. 2b. Also, let the sphere be cut at height h, such that the radius at the intersection of the two cuts will be $\sqrt{r^2 - h^2}$. We then derive the volume of the restricted spherical segment or solid above the given plane by:

$$V_{RSS} = \pi \int_h^r (\sqrt{r^2 - y^2})^2 \, dy$$

$$= \pi \int_h^r (r^2 - y^2) \, dy$$

$$= \pi \left[ r^2 h - \frac{y^3}{3} \right]_h^r$$

$$= \pi \left[ \left[ r^3 - \frac{r^3}{3} \right] - \left[ r^2 h - \frac{h^3}{3} \right] \right]$$

$$= \frac{1}{3} \pi \, (2r^3 - 3r^2h + h^3). \tag{6}$$

The equations $V_{SC}$ and $V_{RSS}$ are basic, intended as guide lines which can be adapted to find a specific volume at a specific location on an individual world.

Equations (2) and (3) are heuristic equations intended for use in planetary science. Notice the insertion of 'passive' in the equations for the Goldilocks Edge; this is an important distinction. The Goldilocks Edge is a region on an astronomical body or planemo where liquid solvents, SPONCH elements, and energy sources exist. However, this in itself does not necessarily lead to the emergence of life, as it is merely the presence of these three factors. It is only when liquid solvents, SPONCH elements and energy sources becomes part of a great prebiotic spot that the possibility of the emergence of life arises. An active definition can thus be derived from this: *The Goldilocks Edge is a spatial and temporal window on an astronomical body or planemo, wherein a great prebiotic spot can exist.*

As discussed in section 2.3, a great prebiotic spot is an active region, a dynamic interaction between liquid solvents, SPONCH elements, and energy sources, where chemical reactions take place, organics are synthesized, useful compounds are selected, and monomers are polymerized. Thus, the Goldilocks Edge is a region on an astronomical body or planemo where a great prebiotic spot can exist at a particular instant in time. Thus, the active Goldilocks Edge can be expressed by the following:

$$G_{active} = \left[\frac{\left(\frac{Z}{L_S} + \frac{E}{L_S}\right)}{V_{SC}}\right]k = \left[\frac{\left(\frac{Z+E}{L_S}\right)}{V_{SC}}\right]k = \left[\frac{C_{RP}}{V_{SC}}\right]k \tag{7}$$

or

$$G_{active} = \left[\frac{\left(\frac{Z}{L_S} + \frac{E}{L_S}\right)}{V_{RSS}}\right]k = \left[\frac{\left(\frac{Z+E}{L_S}\right)}{V_{RSS}}\right]k = \left[\frac{C_{RP}}{V_{RSS}}\right]k \tag{8}$$

where $C_{RP}$ describes the chemical reactions taking place and k is the variability factor or the deterministic reduction coefficient K. For simplicity, molality has been applied in order to better handle the elements and the liquid solvent.

A great prebiotic spot will require protection from the surrounding extreme environment, and a restricted spherical segment could be considered as a good approximation for such protection, as the top layer over the solid could act as a protecting factor. On icy bodies such as Europa the layer of ice at the surface protects the underlying ocean. For Earth-like worlds the atmosphere can presumably act as a protecting layer, if this is considered part of the spherical cap.

The Goldilocks Edge does not disregard the insights of the habitable zone. On the contrary. The spatial and temporal window that was present on the early Earth is evidently due to the Earth's relationship to the sun. The presence of liquid water was because of the distance from the Earth to the sun, and the great prebiotic spot was driven by the radiant energy it received from the sun (although other energy sources such as geothermal energy could also have contributed). Thus, the use of the Goldilocks Edge encompasses the concept of a habitable zone.

What is clarified here however is that although the specific spatial and temporal window on Earth is indebted to the sun, the sun is however not generally the primary source for the spatial and temporal windows. Europa and Enceladus both possess a liquid ocean, and both possess energy sources. They therefore possess a Goldilocks Edge that is not dependent on the sun. In fact, worlds with an ocean protected by an ice cape appear to have a greater chance for the emergence of life than the early Earth on such a foundation, with its exposed and violent surface.

It could be objected that an edge is too restrictive, that feedback from the environment of an entire world is necessary for the emergence of life. However, life itself can indicate that this is not necessarily the case. It is thought that the surface of the Earth once froze entirely, during an event known as the Snowball Earth [Hoffman et al., 1998]. For a time the Earth resembled Europa, with a thick ice cap which shielded the liquid ocean from the radiant energy of the Sun. Liquid water was still present in small amounts. However, water in itself does not achieve much. In fact water presents some obstacles to life, as the chemical reactivity of this solvent leads to the degradation of RNA and DNA, forcing life to make continual repairs to these polymers [Ward and Benner, 2007]. Therefore although the region with liquid water was still quite large, the energy sources has been restricted, cut off from the energy input from the sun as the environment had become.

Yet life was capable of surviving and to emerge again, when the Earth became warmer. Some life may have survived in a dormant state. But some life survived in an active state. The survival and return of these life forms must have taken place in a region, an edge, with not only liquid water, but also with localized energy sources; a spot in the middle of a frozen world.

Thus, to summarize the entire discussion, the following definition can be given: *The active Goldilocks Edge is a spatial and temporal window, a region, on an astronomical body or planemo, wherein a great prebiotic spot can exist; which increase the probability for the emergence of the supra-molecular collection of matter far from thermodynamic equilibrium, which acquires information and exchanges materials and energy with its surrounding abiotic environment, that is, life.*

## 4. Conclusion

Just as it is now known that life can exist and thrive as extremophiles in an extreme edge on a relaxed world, the available evidence likewise suggests that life can emerge in a relaxed edge on an extreme world. Thus, instead of operating within a habitability zone where the focus is on the distance from world to star, it is suggested in this paper that it is better to operate with a Goldilocks Edge where the focus is on a particular region and time on a world, in which life can have the possibility of emerging.

Thus, although the item in the middle, which Goldilocks prefers out of the three items is deemed "just right," this does not mean that the other items, the other worlds that appear to be too extreme; too hot or too cold, too large or too small, in a restricted region within a particular temporal window, do not have the right conditions for chemical evolution. The concept of the Goldilocks zone is focused too narrowly on a world as a whole. Life emerged on an otherwise extreme Earth, seemly unsuited for life. It emerged according to the hypothesis put forward in this paper, in a spatial and temporal window which allowed for the presence of a great prebiotic spot. Such great prebiotic spots on otherwise extreme worlds can exist or have existed in many places in the Solar system on worlds that are considered overall as inhospitable to life.

The habitable zone is thus just one value in the habitability distribution, and considering only this does not allow us to see the whole picture. Such spatial and temporal windows, and the great prebiotic spots within them, can collectively represent the full distribution of possibilities for life in a solar system. This edge, with the right conditions, is not the exception but the rule for the emergence of life. Thus, the Goldilocks Edge predicts that the majority of inhabited worlds in the galaxy are outside the habitable zone.

The concept of a passive Goldilocks Edge is supported by the fact that liquid water can be found on moons such as Europa and Enceladus, and that this liquid water does not occur because of the distance to the sun, but is instead caused by tidal heating. Energy is therefore available on such bodies. The SPONCH elements are widely distributed around the galaxy, and presumably also occur in large amounts on such moons as well. This means that two out of three worlds with the right conditions for life are known to exist outside the habitable zone.

If such a Goldilocks Edge possesses a trend toward becoming an active Goldilocks Edge, with the emergence of a great prebiotic spot, then we would expect to see an increasing trend of such great prebiotic spots over time on astronomical bodies or planemos; where liquid solvents, SPONCH elements, and energy sources exist. If the conjecture is made that life has actually emerged on Europa, Enceladus, and Mars it becomes apparent that 75 % of inhabited worlds exist outside the habitable zone, while only 25 % of the known habitable worlds occur within the habitable zone in this Solar system. Thus, inhabited words outside the zone could dominate in life-bearing solar systems, and the inhabitation of Earth-like planets only appears to be more primary due to sampling bias.

As mentioned in the introduction, it is now well known that there are more planets than stars in the galaxy, and it is already known that there are more planets outside the habitable zone than there are within it. If our Solar system is any indication of the situation in systems further afield, then it follows that there are presumably far more moons than planets in the galaxy (and the number of rogue planets may be very high as well). Thus, it could be the case that although life with a certain probability will emerge on Earth-like planets in the habitable zone because these planets fulfill the conditions for the development of a great prebiotic spot, they may nonetheless still belong to the minority of inhabited worlds in a given Solar system.

This will be in accordance with the Copernican principle, which conjecture that the Earth is not privileged within the universe, and is also in accordance with the Darwinian principle which not only states that single species are not privileged in nature, but also could conjecture, that life on Earth-like planets is not privileged within a Solar system. So while calling it a Goldilocks Edge rather than a Goldilocks zone seems to imply a narrowing down of the possibility of life in a Solar system, it is actually the other way around.

The great prebiotic spot is a novel suggestion from which a number of new questions arise. This paper has mainly focused on the energetic and the evolutionary conditions that life requires. However, there are a number of questions that need answering, regarding the purely physical parameters. Several suggestions have been made as to where such a subglobal prebiotic reactor could be, and thus what it could physically be. Suggestions include lagoons, where tides lead to significant changes in the concentrations of chemical constituents [Nelson et al., 2001], or the hydrothermal zones seen around the oceanic [Baross and Hoffman, 1985]. It could be terrestrial hot springs [Mulkidjanian et al., 2012]. It could be niches such as local ice caps which could additionally provide adequate temperatures for specific chemical reactions to occur.

There are many suggestions, and much debate regarding all of these issues, and no overall consensus has as yet been reached. Such locations could indeed be the home for a great prebiotic spot, and some such locations could provide partial elements for it. However, considering the abundance of proposals and the lack of agreement, these might be insufficient as the physical frames for a great prebiotic spot. A great prebiotic spot may turn out to be something not yet considered, some kind of interdependent buffering systems perhaps.

Another important question that needs to be answered is what size an active Goldilocks Edge must be. Even if the factor k is ignored, and it is assumed that chemical evolution takes place teleologically, what space would be required in order for it to arise? This is not an easy matter to settle. On the Earth biological evolution has long ago taken over, and prebiotic processes cannot compete with biological evolution. Thus, these questions must either be addressed through models, experiments in the laboratory, or through research carried out on other worlds. If life has not already emerged on Enceladus, for example, then a great prebiotic spot may be happening there this very moment, offering rich opportunities for the study of how large such an active Goldilocks Edge must be, for such a *limbus mundi*.